\documentclass[sigplan,screen,nonacm]{acmart}

\newif\ifcameraready%
\camerareadytrue%

\usepackage[capitalise]{cleveref}
\usepackage{svg}
\usepackage{xspace}
\usepackage{listings}
\usepackage{booktabs, multirow}
\usepackage{amsthm}
\usepackage{numprint}
\usepackage{pgfplots}
\usepackage{xcolor}
\usepackage{tabularx}
\usepackage{booktabs}
\usepackage{amsmath}
\usepackage{pifont}
\usepackage{graphicx}
\usepackage{subcaption}
\usepackage{enumitem}
\pgfplotsset{compat=1.18}

\usepgfplotslibrary{groupplots}
\usetikzlibrary{calc}

\definecolor{colorblind_blue}{HTML}{648FFF}
\definecolor{colorblind_purple}{HTML}{785EF0}
\definecolor{colorblind_pink}{HTML}{DC267F}
\definecolor{colorblind_orange}{HTML}{FE6100}
\definecolor{colorblind_yellow}{HTML}{FFB000}
\definecolor{colorblind_green}{HTML}{009E73}

\theoremstyle{definition}
\newtheorem{definition}{Definition}

\newcommand{\name}[1]{\textsc{#1}}
\newcommand{\toolname}{\textsc{zkSBOM}\xspace}

\newcommand{\softwaresupplier}{software supplier\xspace}

\newcommand{\SoftwareSupplier}{Software Supplier\xspace}

\newcommand{\sbomGenerator}{SBOM Generator\xspace}
\newcommand{\sbomgenerator}{SBOM generator\xspace}

\newcommand{\sbomconsumer}{software consumer\xspace}

\newcommand{\SbomConsumer}{Software Consumer\xspace}

\newcommand{\zksbomoperator}{\toolname operator\xspace}
\newcommand{\zksbomOperator}{\toolname Operator\xspace}

\newcommand{\zksbomverifier}{\toolname verifier\xspace}

\newcommand{\securitydb}{vulnerability database\xspace}

\newcommand{\SecurityDb}{Vulnerability Database\xspace}

\newcommand{\tightpar}[1]{{\vspace{2pt}\noindent\bf #1. }}

\setcopyright{acmlicensed}
\copyrightyear{2018}
\acmYear{2018}
\acmDOI{XXXXXXX.XXXXXXX}
\acmConference[Conference acronym 'XX]{Make sure to enter the correct
  conference title from your rights confirmation email}{June 03--05,
  2018}{Woodstock, NY}

\acmISBN{978-1-4503-XXXX-X/2018/06}

\setcopyright{none}               %
\renewcommand\footnotetextcopyrightpermission[1]{} %
\begin{document}

\title[\textsc{zkSBOM}: Privacy-Preserving SBOM Sharing with Zero-Knowledge Sets]{\toolname: Privacy-Preserving SBOM Sharing \\ with Zero-Knowledge Sets}

\ifcameraready%

\author{Tom Sorger}
\email{sorger@kth.se}
\affiliation{%
  \institution{KTH Royal Institute of Technology}
  \city{Stockholm}
  \country{Sweden}
}

\author{Eric Cornelissen}
\email{ericco@kth.se}
\affiliation{%
  \institution{KTH Royal Institute of Technology}
  \city{Stockholm}
  \country{Sweden}
}

\author{Aman Sharma}
\email{amansha@kth.se}
\affiliation{%
  \institution{KTH Royal Institute of Technology}
  \city{Stockholm}
  \country{Sweden}
}

\author{Javier Ron}
\email{javierro@kth.se}
\affiliation{%
  \institution{KTH Royal Institute of Technology}
  \city{Stockholm}
  \country{Sweden}
}

\author{Musard Balliu}
\email{musard@kth.se}
\affiliation{%
  \institution{KTH Royal Institute of Technology}
  \city{Stockholm}
  \country{Sweden}
}

\author{Martin Monperrus}
\email{monperrus@kth.se}
\affiliation{%
  \institution{KTH Royal Institute of Technology}
  \city{Stockholm}
  \country{Sweden}
}

\renewcommand{\shortauthors}{Sorger et al.}

\else

\author{Anonymous author(s)}

\renewcommand{\shortauthors}{Anonymous Authors}

\fi

\begin{abstract}
Software Bills of Materials (SBOMs) are increasingly mandated by regulators, yet existing sharing mechanisms impose a binary choice between full disclosure and full opacity.
This exposes software suppliers to attacks that can be deduced from the SBOM only, such as the presence of a vulnerable dependency.
Conversely, software consumers can be fooled by software suppliers who modify or misrepresent published SBOMs.
We present zkSBOM, a privacy-preserving SBOM sharing mechanism designed to address these threats.
zkSBOM uses zero-knowledge sets to cryptographically commit to the components within an SBOM.
Software consumers can query for known vulnerabilities and receive a cryptographic proof confirming whether the artifact described by the SBOM is affected, without revealing any additional SBOM content.
We conduct a security analysis of zkSBOM by quantifying expected leakage from inclusion and exclusion proofs. We demonstrate real-world feasibility by applying it to realistic scenarios and evaluating its operation requirements.
Our evaluation demonstrates that zkSBOM is a strong, secure, and privacy-preserving mechanism for SBOM sharing, protecting software suppliers and software consumers from one another.
\end{abstract}



\maketitle

\section{Introduction} \label{section:introduction}

Software Bill of Materials, also known as SBOMs, provide structured, machine-readable inventories of all components contained in a software artifact, along with their metadata~\cite{ntia-sbom-sharing}. They are essential elements to compliance mandates, transparency, contractual agreements, and vulnerability management.
Since SBOMs expose the internal architecture of a software artifact, they consequently expose any related vulnerabilities and are highly sensitive to disclosure.

In other words, disclosing full SBOMs is a double-edged sword: it enables full transparency, but at the same time gives all information to potential attackers at no cost, allowing them to weaponize any known vulnerable component~\cite{deng2025sbom}.
This creates substantial tension between transparency and confidentiality in SBOM sharing.

This tension is not just hypothetical.
Berkeley's SBOM Escrow project observes that software vendors resist sharing SBOMs because doing so can disclose closely held proprietary information~\cite{fueyo2023sbom-escrow}.
Cisco's SBOM request workflow is clear: SBOMs are distributed through a customer request process and are explicitly treated as ``proprietary and confidential,'' with restrictions on further sharing~\cite{cisco-sbom-form}.
Deng et al.~\cite{deng2025sbom} empirically demonstrated that even a minimally detailed, regulation-compliant SBOM can reduce adversarial effort to develop working exploits, achieving a 77\% exploit success rate in a controlled setting.
This clearly calls for research on SBOM sharing beyond the naive disclosure of full SBOMS.

In this paper, we address the problem of confidential SBOM sharing with a novel privacy-preserving approach based on \emph{zero-knowledge sets} (ZKS).
ZKS are designed for proving set (non-)membership without revealing any additional information about the set~\cite{micali2003zero}, hence directly applicable to a set of dependencies.
We design, implement, and evaluate \toolname, a system that allows an authenticated \sbomconsumer to query a \softwaresupplier's SBOM and obtain a cryptographic proof of whether it contains components affected by a given vulnerability, without disclosing other SBOM elements.
This provides \sbomconsumer{s} with strong, verifiable assurances about the presence/absence of vulnerable software components, while allowing \softwaresupplier{s} to retain
confidentiality of their
dependency graphs.

We conduct a principled security analysis of zkSBOM against a complete threat model, showing that it satisfies confidentiality, integrity, non-repudiation, and non-equivocation for both the software supplier and software consumer. We characterize the information leakage inherent to SBOM sharing over dependency trees: even when individual queries reveal the presence or absence of a component, public ecosystem metadata such as transitive dependencies, peer relationships, and unique ancestors can allow an adversary to infer additional components. We derive closed-form leakage estimates for both inclusion and non-inclusion proofs.

We experimentally evaluate zkSBOM across three dimensions. For feasibility, we run the zkSBOM on four real-world open-source projects spanning the Cargo, Go, Maven, and npm ecosystems, successfully generating and verifying both inclusion and non-inclusion proofs. For performance, we measure commitment generation, proof construction, and verification times on synthetic SBOMs and over \numprint{43000} real-world SBOMs, finding that all operations complete in well under one second. For leakage, we instantiate our analytical model with empirical dependency data from the top \numprint{10000} packages per ecosystem, finding that non-inclusion proofs leak fewer than two additional components on average.

To sum up, this paper makes the following contributions:
\begin{itemize}
  \item \emph{Novel protocol.}
    We present the first SBOM sharing protocol based on zero-knowledge sets, enabling \sbomconsumer{s} to verify the presence or absence of a specific vulnerability through cryptographic proofs, without directly learning anything about the remaining components (\cref{sec:zksbom-design}).
    This approach dramatically reduces the information available to potential adversaries compared to existing disclosure approaches.
  \item \emph{Security analysis.}
    We provide a security analysis of \toolname to show it provides confidentiality, integrity, non-repudiation, and non-equivocation (\cref{section:security-analysis}).
    We also quantify information leakage arising from privacy-preserving proofs given knowledge of public data.
  \item \emph{Practical feasibility results.}
    We demonstrate that \toolname works end-to-end across four software ecosystems
    with runtime and storage overheads that meet operational requirements of all stakeholders (\cref{section:evaluation}).
  \item \emph{Prototype implementation.}
    We share a prototype implementation of \toolname (\cref{section:implementation}) publicly on GitHub~\cite{zksbom} for facilitating future research on this important topic.
\end{itemize}

\begin{figure*}[t]
    \centering
    \includegraphics[width=0.70\textwidth]{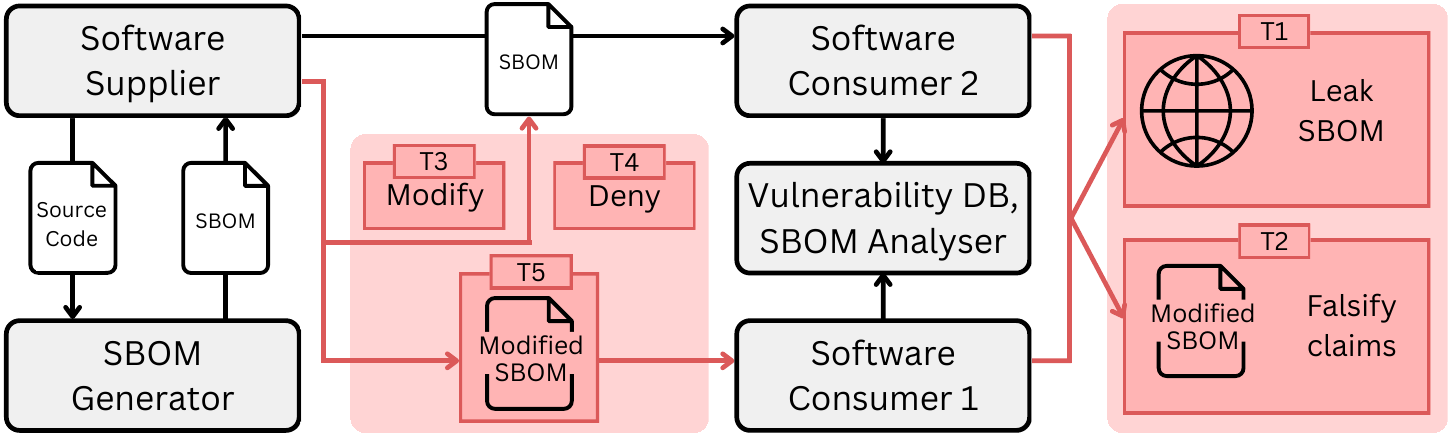}
    \caption{Overview of the problem statement.}
    \label{fig:problem-statement}
\end{figure*}

\section{Background} \label{section:background}
This section provides the necessary background on SBOMs, zero-knowledge proof systems, and cryptographic primitives used.

\subsection{SBOM}
A \emph{Software Bill of Materials} (SBOM) is a structured, machine-readable inventory of all components contained in a software artifact, along with metadata describing each component~\cite{ntia-sbom-define}.
Each component entry can include identifiers such as package name, version, supplier, and ecosystem, as well as provenance metadata such as cryptographic hashes and licensing information.
Several standard formats exist for encoding SBOMs, including CycloneDX~\cite{cyclonedx}, SPDX~\cite{spdx}, and SWID Tags~\cite{swid}.
For the purpose of this work, we are only interested in the component parts in SBOMs---the collection of third-party packages included in a software artifact.

SBOMs are increasingly mandated by regulations; for example, the EU Cyber Resilience Act (CRA)~\cite{eu-cra} requires manufacturers of products with digital elements to produce and maintain a machine-readable SBOM covering at least top-level dependencies.
Notably, manufacturers are not obliged to make SBOMs public, yet must be able to submit them to authorities in some form.
In practice, SBOMs are distributed as plain files on public or private repositories, in product documentation, or shared via email~\cite{ntia-sbom-sharing}.

\subsection{Zero-knowledge} \label{sec:zero-knowledge}
A \emph{zero-knowledge proof} (ZKP) is a cryptographic protocol in which a \emph{prover} convinces a \emph{verifier} that a given statement is true without revealing any information beyond the validity of the statement itself~\cite{goldwasser1989knowledge}.
Zero-knowledge proofs satisfy three properties: \emph{completeness} (an honest prover always convinces an honest verifier), \emph{soundness} (a dishonest prover cannot convince the verifier of a false statement, except with negligible probability), and \emph{zero-knowledge} (the verifier learns nothing beyond the truth of the statement).
A proof is \emph{verified} when the verifier, using only the proof and public parameters, confirms that the prover's claim is valid according to the verification algorithm without needing access to the prover's private inputs.

\tightpar{Zero-Knowledge Sets}
A \emph{zero-knowledge set} (ZKS)~\cite{micali2003zero} allows a prover to commit to a finite set $S$ and subsequently prove, for any queried element $x$, whether $x \in S$ or $x \notin S$, without revealing any additional information about $S$, including its size.
The prover first publishes a short \emph{commitment} to the set.
For each query, the prover generates a proof of membership or non-membership that any party can verify against this commitment.

The elements of $S$ can be arbitrary strings obtained by serializing richer data, so long as membership is tested on the same encoding the prover committed to.
For example, one can represent a key-value dataset as a set of encoded pairs, committing to the whole collection while later proving only whether a specific key-value pair appears in it.

\subsection{Definitions} \label{section:definitions}

In this work, we use the following cryptographic schemes:
a hash function $H$,
Digital Signature scheme $DS$~\cite{goldwasser1988digital},
Zero-Knowledge Set scheme $ZKS$~\cite{micali2003zero},
and a transparency log $TL$~\cite{10.1145/3319535.3345652}.
We omit security parameters for brevity and use explicit random inputs, $r$, in our notation (over randomized algorithms).

\begin{definition}
Let $H$ be a cryptographic hash function that maps variable-length input to fixed-length output.
We denote this operation as $h \gets H(m)$.
The function $H$ must be cryptographically secure, namely collision-, preimage-, and second preimage-resistant.
\end{definition}

\begin{definition}
Let $DS$ be a digital signature scheme that consists of algorithms $DS.\mathit{KeyGen}$, $DS.\mathit{Sign}$, and $DS.\mathit{Verify}$:

$(pk, sk) \gets DS.\mathit{KeyGen}(r)$:
This algorithm generates a fresh public ($pk$) and private ($sk$) key pair. %

$\sigma \gets DS.\mathit{Sign}(m, sk)$: 
This algorithm denotes the operation of signing a message $m$ with a private key $sk$ to produce a signature $\sigma$.

$b \gets DS.\mathit{Verify}(\sigma, m, pk)$:
This algorithm verifies that a signature $\sigma$ was generated for message $m$ with the private key corresponding to the public key $pk$, returning a bit $b$ indicating whether this is the case.

The $DS$ scheme must be correct and secure, guaranteeing integrity and unforgeability.
\end{definition}

\begin{definition}
Let $ZKS$ be a Zero-Knowledge Set scheme that consists of the algorithms $ZKS.\mathit{Commit}$, $ZKS.\mathit{Query}$, and $ZKS.\mathit{Verify}$:

$(c, s) \gets ZKS.\mathit{Commit}(D, r)$:
This algorithm takes a datastore $D$ to commit to.
The datastore is a collection of label ($l$) and value ($v$) pairs, $D = \{(l_1, v_1), \cdots, (l_n, v_n)\}$. %
The algorithm produces a commitment $c$ to the datastore and secret state $s$.

$(\pi, v) \gets ZKS.\mathit{Query}(s, D, l, r)$:
This algorithm takes the secret state $s$, datastore $D$, and a label $l$ to query as input, and returns a proof $\pi$ and the value $v$ at the given label (if $\nexists v. \; (l, v) \in D$ then $v = \bot$).

$b \gets ZKS.\mathit{Verify}(c, l, v, \pi)$:
This algorithm takes a commitment $c$, a label-value pair $(l, v)$, and a proof $\pi$ as input and returns a boolean value $b$ indicating whether the proof is correct.

The $ZKS$ scheme must satisfy the properties completeness, soundness, and zero-knowledge.
\end{definition}

\begin{definition}
Let $TL$ be a Transparency Log scheme that consists of the algorithms $TL.\mathit{Setup}$, $TL.\mathit{Append}$, $TL.\mathit{Lookup}$, and $TL.\mathit{Verify}$.
For brevity, we omit from our definition the algorithms related to verifying the append-only property.

$(pp, VK) \gets TL.\mathit{Setup}(r)$:
This algorithm initializes the transparency log and returns the public parameters $pp$ and the verification key $VK$.

$(S_j, d_j) \gets TL.\mathit{Append}(pp, S_i, d_i, k)$:
This algorithm appends the value $k$ to the transparency log state $S_i$.
It creates the next state, $j=i+1$, and corresponding digest.

$(b, \pi) \gets TL.\mathit{Lookup}(pp, S_i, k)$:
This algorithm looks up the value $k$ in the state $S_i$, returning whether the value is present as a bit $b$ and a proof of (non-)inclusion $\pi$.

$b \gets TL.\mathit{Verify}(VK, d_i, k, b, \pi)$:
This algorithm takes a proof $\pi$ for (non-)inclusion of $k$,  in accordance with $b$, in the transparency log at the state associated with the digest $d_i$.

The $TL$ scheme must guarantee the correctness and security of both membership and append-only.
\end{definition}

Additionally, we assume three helper functions.
The function $\mathit{ToDatastore}$ converts SBOMs to data stores.
We denote by $D \gets \mathit{ToDatastore}(x)$ the process of extracting data from structured input $x$ to get a datastore $D$.
The pair $\mathit{Store}$ and $\mathit{Load}$ can be used to store and load data locally. $\bot \gets \mathit{Store}(k, v)$ stores value $v$ under key $k$ and $v \gets \mathit{Load}(k)$ loads the value previously stored for $k$.

\section{Problem Statement and Threat Model} \label{section:problem-statement}

Current SBOM sharing practices offer an all-or-nothing solution between full disclosure and full concealment, leaving neither suppliers nor consumers adequately served.
This section motivates the problem through a concrete scenario and then presents the actors, security goals, and threat model.

\subsection{Problem Statement}\label{sec:problem-statement}
SBOMs are distributed via public repositories (e.g., GitHub), shared via email~\cite{painchek-sbom}, or published as a URL in product documentation~\cite{ntia-sbom-sharing}.
When received, the \sbomconsumer gains unrestricted access to the full SBOM.
This raises confidentiality concerns for the software supplier, because a detailed SBOM may expose proprietary architectural decisions and dependency information~\cite{cisco-sbom-form}, which can in turn facilitate targeted attacks.
On the other hand, the software consumer has no control over the trustworthiness of the SBOM shared by the software supplier.

For instance, consider a hospital IT operator running a vendor-supplied clinical information system on the morning Log4Shell is disclosed.
The operator must urgently decide how to react: either disconnect the system and wait for a vendor patch, or continue running the system at risk.
This decision depends on a single bit of information: \emph{is the vendor's product affected?}
The trade-off is non-trivial, reacting without this information can expose sensitive health data if the artifact is vulnerable, or needlessly disrupt critical care if it is not.
To take this decision, the operator does not need to learn the rest of the dependency graph, and the vendor may have commercial reasons not to share it.
At the same time, the operator should not accept the vendor's claim at face value: the same commercial reasons give the vendor an incentive to downplay exposure during the public-relations window.
Scenarios like this motivate the need for an SBOM sharing system that yields a verifiable, cryptographically backed \emph{yes}-or-\emph{no} statement about SBOM contents, with no further information disclosure.

\subsection{Security Goals and Threat Model}\label{sec:threats}
We characterize an SBOM sharing system along three dimensions: the actors involved, the security goals it must uphold, and the threats it must mitigate.
Our threat model focuses on adversarial behavior between the \softwaresupplier and \sbomconsumer, who have conflicting incentives around disclosure.
We consider threats originating from an \sbomgenerator to be out of scope.

\tightpar{Actors}
There are three main actors in a SBOM sharing approach: the \softwaresupplier, \sbomconsumer, and \sbomgenerator.
The \emph{\SbomConsumer} is the actor that wants to use the SBOM and is concerned with the trustworthiness of the SBOM.
This can be, for example, a user of the software, a government agency, or a compliance organization.
The \emph{\SoftwareSupplier} is the actor that develops and distributes proprietary software artifacts and is concerned with the confidentiality of the SBOM.
The \emph{\sbomGenerator} is a third-party software service that analyses the \softwaresupplier's software artifact and generates an SBOM.

In addition to the main actors, \textit{\SecurityDb{s}} play a key role by enabling the retrieval of vulnerability information based on specific identifiers (e.g., CVEs).

\tightpar{Security Goals}
We outline four security goals for an SBOM sharing system in a setting of a privacy relationship with mutual distrust between the supplier and the consumer.

\emph{Confidentiality (G1)} requires the system to ensure that sensitive SBOM data is accessible only to authorized entities. This includes selective disclosure of SBOM dependencies, enabling \sbomconsumer{s} to verify security claims without \softwaresupplier having to expose the full SBOM.
It also requires that \softwaresupplier has the ability to revoke access, e.g., if the business relationship with a \sbomconsumer terminates.
\emph{Integrity (G2)} requires that SBOMs remain provably unaltered after release by \softwaresupplier. Modification from \softwaresupplier must be detectable by \sbomconsumer and vice versa.
\emph{Non-Repudiation (G3)} requires the system to ensure that \softwaresupplier{s} cannot retroactively deny the authorship or existence of their published SBOMs.
Finally, \emph{Non-Equivocation (G4)} requires the system to guarantee that all \sbomconsumer{s} receive the same SBOM for a specific software artifact.

\tightpar{\SbomConsumer Threats}
A malicious \sbomconsumer can compromise three security goals.
They can compromise confidentiality (G1) by \emph{leaking the SBOM to an unauthorized third party or by uploading it to an insecure or public endpoint (T1)}, thereby violating the supplier's intellectual property rights.
This includes keeping a copy of the SBOM after their business contract ends, allowing  \sbomconsumer{s} to continue scanning the SBOM for new zero-day vulnerabilities to exploit.
A malicious \sbomconsumer can compromise integrity (G2) by \emph{altering the received SBOM, making false claims, and accusing \softwaresupplier of providing a false SBOM (T2)}.

\tightpar{\SoftwareSupplier Threats}
A malicious \softwaresupplier can break three security goals.
First, they can break integrity (G2) to, after a new vulnerability is published, \softwaresupplier \emph{retroactively modify the released SBOM (T3)} by excluding or hiding vulnerable components.
Second, a malicious \softwaresupplier can break non-repudiation (G3) by sharing an SBOM with \sbomconsumer and later \emph{claiming it is not an official SBOM for their software artifact (T4)}.
Finally, a malicious \softwaresupplier can break non-equivocation (G4) if  \softwaresupplier \emph{provides different SBOMs to different \sbomconsumer{s} (T5)}, such as sharing the full SBOM with regulators while providing a fake, safe version to business customers.

\section{Design of \toolname} \label{sec:zksbom-design}
This section provides an overview of \toolname, our proposed privacy-preserving SBOM sharing system for protecting against the threats described in Section \ref{sec:threats}. \toolname's architecture enables selective disclosure of any SBOM data. Our primary use case targets the selective disclosure of vulnerable components to \sbomconsumer{s}.

\subsection{System Overview} \label{sec:system-overview}
We introduce the role of \emph{\zksbomOperator}.
A dedicated system responsible for answering privacy-preserving vulnerability queries from \sbomconsumer{s} without exposing the full SBOM contents of the \softwaresupplier.

First, the \softwaresupplier generates a comprehensive SBOM of the software artifact (\Cref{sec:sbom-generation}).
Second, the \softwaresupplier uploads the SBOM to the \zksbomoperator, which constructs a cryptographic commitment of its components (\Cref{sec:commitment-creation}).
Third, the \softwaresupplier verifies the commitment's accuracy and publishes it, along with metadata, to a transparency log (\Cref{sec:verify-commitment-correctness}).
It then distributes the software artifact to the \sbomconsumer.
Fourth, an authorized \sbomconsumer can submit a vulnerability query to the \zksbomoperator (\Cref{sec:consumer-query}).
Fifth, the \zksbomoperator generates a zero-knowledge proof confirming the presence or absence of the queried vulnerability from the SBOM and returns it (\Cref{sec:proof-generation}).
Sixth, the \sbomconsumer verifies the correctness of the received proof (\Cref{sec:verify-proofs}).

\begin{figure*}[tbp]
    \centering
    \includegraphics[width=0.85\textwidth]{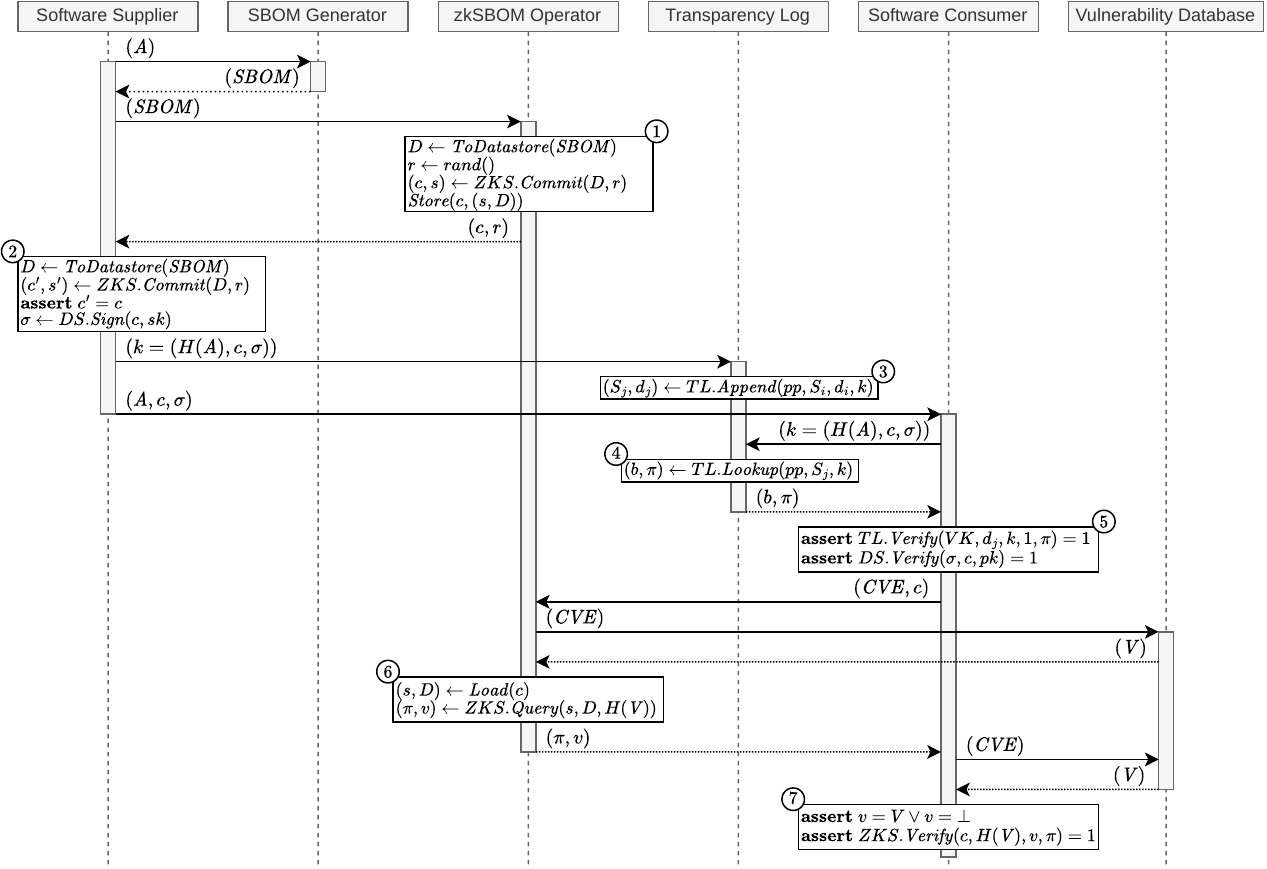}
    \caption{Sequence diagram of the \toolname protocol that provides privacy guarantees in the context of SBOM sharing, while allowing verifiable querying of vulnerability presence.}
    \label{fig:protocol}
\end{figure*}

\subsection{Protocol} \label{sec:protocol}
\Cref{fig:protocol} provides an overview of the protocol underpinning \toolname.
Our presentation of the protocol assumes all cryptographic schemes defined in \Cref{section:definitions} are appropriately initialized with keys distributed to relevant actors.
We furthermore assume every interaction is appropriately authenticated so that, for example, when the \sbomconsumer queries the \zksbomoperator, it can be determined that the \sbomconsumer is allowed to query the \softwaresupplier's artifact.

\subsubsection{SBOM Generation} \label{sec:sbom-generation}
Upon completion of the software development phase, the \softwaresupplier has to generate an SBOM.
This process is performed by an \sbomgenerator, which extracts dependencies, components, and metadata from the software artifact $A$.
Once the generation is complete, the \softwaresupplier reviews and submits the SBOM to the \zksbomoperator.

\subsubsection{Commitment Creation} \label{sec:commitment-creation}
After receiving a submission, the \zksbomoperator, as seen in box~(1) of \Cref{fig:protocol},
converts the SBOM into a datastore $D$ and generates a fresh random seed $r$ to construct a zero-knowledge set, internally represented by $s$, and corresponding commitment $c$.
The commitment uniquely represents the \softwaresupplier's provided SBOM.

The datastore consists solely of the dependencies $d_1, \ldots, d_n$ listed in the SBOM, stored as $(H(d_i), d_i)$ key-value pairs, and does not contain any data that may change over time (e.g., vulnerability information).
A new commitment has to be created for every new version of the software artifact.

Finally, the commitment $c$ and the random seed $r$ are returned to the \softwaresupplier.

\subsubsection{Commitment Publication} \label{sec:verify-commitment-correctness}
After receiving the commitment and associated random seed, the \softwaresupplier, as seen in box~(2) of \Cref{fig:protocol}, verifies the accuracy of the generated commitment $c$ as computed by the \zksbomoperator.
To do this, it repeats the computation using the provided random seed $r$ and compares the resulting commitment $c'$.
A matching result guarantees that the \zksbomoperator did not commit to a tampered SBOM.

Once the commitment is validated, the \softwaresupplier uses its private key $sk$ to sign the commitment $c$, obtaining $\sigma$.
It then publishes the commitment $c$, signature $\sigma$, and artifact hash $H(A)$ to a public transparency log, box~(3) in \Cref{fig:protocol}.
The append-only property of the transparency log can be used to ensure only one commitment is published per $H(A)$ (by the \softwaresupplier). This ensures all \sbomconsumer{s} agree on the commitment and hence SBOM.

After that, the \softwaresupplier can deliver the software artifact to the \sbomconsumer.

\subsubsection{\SbomConsumer Query} \label{sec:consumer-query}
After obtaining the software artifact $A$, commitment $c$, and signature $\sigma$, the \sbomconsumer computes $H(A)$ and queries the transparency log for its inclusion.
The transparency log will produce a proof of (non-)inclusion of the triple.
The \sbomconsumer checks the transparency log's inclusion proof $\pi$ and \softwaresupplier's signature $\sigma$ over the commitment $c$ using its public key $pk$, as seen in box~(5) of \Cref{fig:protocol}.

Using the verified commitment, the \sbomconsumer is empowered to ask the \zksbomoperator whether the software artifact it is using is affected by a known vulnerability, so that it may, e.g., implement appropriate defenses.
To perform such a query, the \sbomconsumer looks up the identifier associated with the vulnerability (e.g., a CVE) and sends it, along with the commitment $c$, to the \zksbomoperator who can determine if the SBOM associated with $c$ contains the component(s) associated with the CVE.

\subsubsection{Proof Generation} \label{sec:proof-generation}
Upon receiving a query request, the \zksbomoperator queries the \securitydb for information about the component $V$ associated with the provided vulnerability identifier.
Then the \zksbomoperator, as seen in box~(6) of \Cref{fig:protocol}, loads the data associated with the given commitment $c$, computes $l = H(V)$ and generates a (non-)inclusion proof $\pi$.
If the component is in the SBOM (i.e., $(l, V) \in D$), this will be a proof of inclusion. Otherwise, if $(l, V) \notin D$, this will be a proof of non-inclusion.

This cryptographic proof demonstrates whether the vulnerable component exists in the committed SBOM, without revealing any other information.
In case of a non-inclusion proof, this provides a verifiable guarantee that the artifact is not affected by the queried vulnerability.
If the vulnerability affects multiple versions of a component, an inclusion or non-inclusion proof is generated for all versions.
If the vulnerability affects multiple components, this is repeated for each component.

The proof $\pi$ and datastore value $v$ are returned to the \sbomconsumer for verification.

\subsubsection{Proof Verification} \label{sec:verify-proofs}
After receiving the proof $\pi$ for component $v$ from the \zksbomoperator the \sbomconsumer, as seen in box~(7) of \Cref{fig:protocol}, performs a local verification to validate the claim.
To do this, it queries the \securitydb for the affected component $V$ and asserts $v = V \lor v = \bot$, ensuring the proof is either a contextually relevant inclusion proof or a non-inclusion proof.
If so, the \sbomconsumer uses the $ZKS.\mathit{Verify}$ algorithm to cryptographically verify the proof.

As a result, the \sbomconsumer knows with certainty whether the artifact it is using is affected by the vulnerability of interest without having learned anything about the SBOM beyond that fact.

\section{Security Analysis}\label{section:security-analysis}
This section presents a security analysis of the \toolname system w.r.t. our threat model (\Cref{sec:threats}).
We do this by considering, in turn, what might happen if each of the actors were malicious.
Recall that we do not consider a malicious \sbomgenerator or \securitydb in our threat model.

\subsection{\SbomConsumer}
We assume an adversarial \sbomconsumer  that only has access to the software artifact, \zksbomoperator, \securitydb, and transparency log. Moreover, we assume it is not possible to determine components used from the software artifact.
We grant this adversary perfect knowledge about how frequently components are being used through download statistics. This enables the adversary to determine unique dependency relations.
Namely, consider a component A which depends on a component B: if both A and B have exactly 100 downloads, it must be true (under this assumption) that A is the only component depending on B.

Our analysis will focus on the knowledge the adversary gains with certainty. We also assume that all SBOM components are used by the \softwaresupplier's artifact.

\subsubsection{Confidentiality (T1)}
A malicious \sbomconsumer wants to break the confidentiality guarantees of the \softwaresupplier's SBOM. The use of ZKS theoretically prevents the \sbomconsumer from learning anything about the SBOM beyond the known vulnerable components it contains, and thus prevents leaking more information.
However, this information does not exist in a vacuum. The context of a public packaging ecosystem can reveal further details about the composition of the target artifact.

In particular, in public software ecosystems, the \emph{relations} between components are known. These relations can be used to infer the presence of other components in the target artifact beyond the query subject(s). There are three types of component relationships that can be used for this purpose: descendant, peer, and ancestor.
We will consider each in turn.

\tightpar{Descendants}
When an inclusion proof demonstrates the use of a component, this leaks information about the descendants (or \emph{transitive dependencies}) of that component. The impact of this leakage depends on the position of the vulnerable component in the dependency tree; leaves leak no additional information, while direct dependencies have the potential to reveal the most.
If, instead, a non-inclusion proof demonstrates that a component is not in use, it leaks that the component's unique descendants are also not in use.

\tightpar{Peers}
Some components are exclusively used alongside other components--e.g., plugins--which we refer to as peers. Consider the case of the component \texttt{express} and plugin \texttt{express-session}.
When \texttt{express-session} is the subject of an inclusion proof, this leaks that \texttt{express} is used too.
If instead \texttt{express} is the subject of a non-inclusion proof, this leaks that \texttt{express-session} also is not in use.

On the other hand, an inclusion proof for \texttt{express} does not necessarily leak usage of \texttt{express-session}, only making its use more probable. Similarly, for a non-inclusion proof for \texttt{express-session} w.r.t. \texttt{express}.

\tightpar{Ancestors}
When a component is known to be used due to an inclusion proof, it reveals its ancestors if it is a unique descendant of that ancestor.
Similarly, a non-inclusion proof for a component leaks a unique ancestor is not in use either.

\paragraph{Leakage Quantification}
Based on this analysis, we propose a novel metric to estimate the number of additional components leaked by an inclusion and a non-inclusion proof. 

Let $E[DC]$ and $E[PC]$ be the expected number of descendant and peer components, respectively.
It could be that $E[PC] < 1$, which intuitively means that, on average, $1$ in $n$ components have a (single) peer.
Let $P[AC]$ be the probability that any given component has a unique ancestor.
The leakage $L$ is the number of \emph{additional} components the attacker can infer when learning a component is, or is not, used.

For an inclusion proof, the leakage, or the number of additional components known to be used, is
\begin{equation*}
    \begin{aligned}
        E[L_i] &= (1-P[AC]) \times (E[DC] + E[PC] \times (1 + E[DC])) \\
               &+ P[AC] \times E[DC] \\
    \end{aligned}
\end{equation*}
Intuitively, if the queried component has multiple ancestors ($1-P[AC]$), it leaks its own descendants ($E[DC]$) as well as any peers and their descendants ($E[PC] \times (1 + E[DC])$). Otherwise, it leaks the ancestor and its descendants ($P[AC] \times E[DC]$), which subsumes the other terms.

For a non-inclusion proof, we define $E[DC_u] = P[AC] \times E[DC]$ (unique descendants) for convenience. Then the leakage, or the number of additional components known not to be used, is
\begin{equation*}
    \begin{aligned}
        E[L_e] &= (1-P[AC]) \times (E[DC_u] + E[PC] \times (1 + E[DC_u])) \\
               &+ P[AC] \times (1 + E[DC_u]) \\
    \end{aligned}
\end{equation*}
Intuitively, if it has multiple ancestors ($1-P[AC]$) it leaks its own unique descendants ($E[DC_u]$) and its peers and their unique descendants ($E[PC] \times (1 + E[DC_u])$) are not used. Otherwise ($P[AC]$), it leaks the ancestor is not used either, along with any other unique dependency ($\times (1 + E[DC_u])$).

We estimate transitive and peer dependency numbers for 4 ecosystems and quantify the resulting leakage in \Cref{section:evaluation}.

\paragraph{Ambiguity}
The described modes of leakage are inherently ambiguous due to version ranges, overrides, and constraints from other, unknown components.
Each of these factors affects the set of possible component-version pairs related to the subject of a proof.
The use of version ranges naturally increases the set from one to many, while constraints from other components can affect the selected version in certain resolution algorithms (e.g., Minimal Version Selection in Go).
Overrides can even push the selected version beyond publicly declared supported versions.
This ambiguity cascades as variability in a direct dependency of a proof's subject means that the transitive resolution of all possible versions is feasible.

However, the ambiguity is reduced by the time at which the SBOM was published (ensuring no newer components are used) and by proofs of SBOMs for earlier versions of the same software (since components revealed earlier are more likely to still be in use).

\subsubsection{Integrity (T2)}
A malicious \sbomconsumer wants to claim the \softwaresupplier's SBOM contains components it does not. The \sbomconsumer is unable to do this because it would require fabricating proofs of inclusion for unused components or proofs of non-inclusion for used components. The use of ZKS and universal agreement on the commitment prevent the creation of such proofs.

\subsection{\SoftwareSupplier}
We assume the \softwaresupplier generates honest SBOMs, even though their incentives are to the contrary (e.g., it wants to omit known vulnerable components). The \softwaresupplier must trust the \sbomgenerator with its source code and the \zksbomoperator with its SBOM.

\subsubsection{Integrity (T3)}
A malicious \softwaresupplier wants to claim its product is unaffected by a disclosed vulnerability in a component present in a committed SBOM.
This attack is prevented because it would require creating a non-inclusion proof for the vulnerable component that is consistent with the commitment, which is impossible with ZKS.

\subsubsection{Non-Repudiation (T4)}
A malicious \softwaresupplier wants to claim an SBOM is illegitimate, e.g., because it contains a known vulnerable component.
This attack is prevented by having the \softwaresupplier sign the SBOM commitment using their private key.

\subsubsection{Non-Equivocation (T5)}
A malicious \softwaresupplier wants to present different components or SBOMs to different \sbomconsumer{s},
also known as a split-view attack.
First, publishing the signed SBOM commitment to a public transparency log ensures that all \sbomconsumer{s} agree on it.
Second, because all \sbomconsumer{s} agree on the commitment, the ZKS prevents the \softwaresupplier from fabricating diverging proofs about components.

\subsection{\zksbomOperator}
As an independent actor, the \zksbomoperator must be trusted not to leak the SBOM outright (confidentiality, G1), which it is disincentivized from doing due to its business interests.
Recall the \zksbomoperator only ever sees the \softwaresupplier's SBOM.

\subsubsection{Integrity (T6)}
The \zksbomoperator cannot create proofs for an SBOM that contains missing or extra components compared to the \softwaresupplier's SBOM.
The \softwaresupplier checks the commitment independently, and the \sbomconsumer uses the \softwaresupplier-provided commitment, of which it verified the signature.%

Being tied to an SBOM the \softwaresupplier agrees with, the \zksbomoperator cannot convince \sbomconsumer{s} the \softwaresupplier uses a component it does not (e.g., to cause reputational harm by pretending a vulnerable component is used) or does not use a component it does (e.g., to prevent a \sbomconsumer from implementing defenses against a known vulnerability).
This is prevented because it is not possible to fabricate such inclusion or non-inclusion proofs that are in agreement with the commitment.

\section{Implementation} \label{section:implementation}
\toolname is implemented as an open-source prototype written in Rust~\cite{zksbom}.
The prototype consists of two components:
the \texttt{zksbom-operator}, corresponding to the \zksbomoperator, which handles the SBOM upload process and proof generation,
and the \texttt{zksbom-verifier}, a lightweight tool designed for the \sbomconsumer to validate the generated inclusion or non-inclusion proofs.
\toolname currently does not implement recomputation of the commitment, signing, and the transparency log.

For the zero-knowledge operations, the prototype uses Microsoft's Ordered Zero-Knowledge Set (oZKS) library~\cite{ozks}.
This library provides the data structures and zero-knowledge primitives needed to prove whether a package is present in or absent from an SBOM.
We use BlakeTwo256 as our hashing algorithm.
The prototype supports the CycloneDX SBOM standard~\cite{cyclonedx} and can be trivially extended to support additional SBOM formats.
To establish a reliable ground truth for the proof generation process, the system relies on the \href{https://github.com/advisories}{GitHub Advisory Database} as the \securitydb.

\section{Experimental Evaluation} \label{section:evaluation}

This section presents the evaluation of \toolname, answering the following request questions:

\begin{enumerate}[label=RQ\arabic*.]
    \item \emph{(Feasibility)} Can \toolname correctly generate and verify (non-)inclusion proofs for real-world software artifacts across diverse package ecosystems?
    \item \emph{(Performance)} Does \toolname perform practically w.r.t. time and memory usage during SBOM upload, proof construction, and proof verification?
    \item \emph{(Leakage)} How much does (non-)inclusion proof from \toolname leak according to the leakage model from \Cref{section:security-analysis} based on real-world ecosystem data?
\end{enumerate}

\subsection{Experimental Methodology}

\subsubsection{RQ1: Feasibility}

To assess whether \toolname is feasible in practice, we run the end-to-end pipeline across four package ecosystems: Cargo, Go, Maven, and npm.
For each ecosystem, we select one popular, open-source project known to contain at least one package with a published CVE as a direct or transitive dependency.
We also select CVE-2025-55182 as the non-inclusion target across all subjects because none of the projects depend on any of the affected components.

We generate a CycloneDX SBOM for the subjects using
\href{https://crates.io/crates/cargo-cyclonedx/0.5.9}{\texttt{cargo-cyclonedx}} for Cargo,
\href{https://pkg.go.dev/github.com/CycloneDX/cyclonedx-gomod@v1.7.0}{\texttt{cyclonedx-gomod}} for Go,\\
\href{https://central.sonatype.com/artifact/org.cyclonedx/cyclonedx-maven-plugin/2.9.1}{\texttt{cyclonedx-maven-plugin}} for Maven, and
\href{https://www.npmjs.com/package/@cyclonedx/cyclonedx-npm/v/4.2.1}{\texttt{cyclonedx-npm}} for npm.
The resulting SBOMs are uploaded to a local \toolname instance, which builds a ZKS commitment over the full dependency sets.
For each subject, we then issue two queries against the committed SBOM:

\tightpar{Inclusion proof}
    We simulate a \sbomconsumer querying whether the software artifact is vulnerable to a CVE known to affect one of its dependencies.
    \toolname resolves the CVE to the corresponding package identifier and produces an inclusion proof, which is successfully verified by \texttt{zksbom-verifier}.

\tightpar{Non-inclusion proof}
    We again simulate a \sbomconsumer querying whether the software artifact is vulnerable to CVE-2025-55182, a vulnerability affecting none of the subjects.
    \toolname resolves the CVE to the affected package identifiers and produces a non-inclusion proof for each, which is successfully verified by \texttt{zksbom-verifier}.

A run is considered \emph{successful} when 1) \toolname terminates without error and 2) \texttt{zksbom-verifier} successfully verifies the proof.
We treat this binary pass-or-fail criterion as our feasibility metric for RQ1.

\subsubsection{RQ2: Performance}

We evaluate the performance of \toolname to assess its practical applicability by measuring execution time and storage overhead.
To enable controlled experiments for both inclusion and non-inclusion proofs, we generate synthetic SBOMs with dependency counts ranging from \numprint{0} to \numprint{1000}.
Each synthetic SBOM is constructed to include a dependency vulnerable to Log4Shell (CVE-2021-44228), which is used to generate inclusion proofs.
Inclusion proofs are omitted for SBOMs with zero dependencies.
For non-inclusion proofs, we use CVE-2026-35613 because it has only one vulnerable component.
Measurements on synthetic SBOMs are repeated ten times, and the median is reported.

We further evaluate the extremes by creating a set of SBOMs with \numprint{1000} components each and a variable number of vulnerable components, ranging from \numprint{1} to \numprint{1000}.
We also create one SBOM with \numprint{1000} components and no vulnerable components. %

Our next experiment considers real-world SBOMs from the Wild SBOMs dataset \cite{soeiro2025wildsbomslargescaledataset}.
To enable a direct comparison with our synthetic SBOMs, we select all SBOMs in the CycloneDX format containing at most \numprint{1000} dependencies.
This filtering yields \numprint{52057} SBOMs, of which \numprint{8117} are excluded due to unexpected formatting and unsupported ecosystems, resulting in a final dataset of \numprint{43940} SBOMs (55.89\% of the original dataset).
Measurements on real-world SBOMs are performed once.
If multiple SBOMs share the same number of components, we report the median across those instances.
The dataset contains no ground truth to indicate which components are vulnerable.
As a result, we cannot meaningfully generate inclusion and non-inclusion proofs for these SBOMs.
Hence, we limit our experiments to synthetic SBOMs, where the presence or absence of vulnerabilities can be controlled.

\tightpar{Timing}
We measure the time required to upload an SBOM to the \zksbomoperator, including parsing and commitment generation (box~(1) in~\Cref{fig:protocol}).
We also measure the time to generate inclusion and non-inclusion proofs, excluding the time required to query the \securitydb (box~(6) in~\Cref{fig:protocol}) and the time required for the \zksbomverifier to validate the proofs (box~(7) in~\Cref{fig:protocol}), both for a variable number of SBOM components and fixed number of one vulnerable component and a fixed number of \numprint{1000} SBOM components and a variable number of vulnerable components.

\tightpar{Storage}
We measure storage overhead as the sizes of the database files generated per SBOM and the corresponding inclusion and non-inclusion proof files, for both a variable number of SBOM components and a fixed number of 1 vulnerable component, and a fixed number of \numprint{1000} SBOM components and a variable number of vulnerable components.

For both timing and storage experiments with a variable number of vulnerable components, the inclusion proofs are performed using CVE-2026-39356, since the number of vulnerable components exceeds \numprint{1000}, by replacing the required number of components with vulnerable ones.
For non-inclusion proofs, we query the \href{https://nvd.nist.gov/}{National Vulnerability Database (NVD)} for the latest \numprint{5000} CVEs and use them in our experiment.
The experiment was conducted on April 6, 2026.

All performance experiments were conducted on a MacBook Pro equipped with an Apple M3 Pro processor (12-core CPU) and 18 GB unified memory, running macOS 26.4.1.

\subsubsection{RQ3: Leakage}
\label{sec:rq3-methodology}
We evaluate actual leakage based on real-world ecosystem data.
We focus on leakage through \emph{transitive} and \emph{peer} components.
In line with RQ1, we consider the Cargo, Go, Maven, and npm ecosystems.
As a representative sample of components used in the real world, this evaluation is limited to the top \numprint{10000} most used components in each ecosystem.
We use ecosyste.ms' packages API~\cite{ecosystems} to obtain a list of the names of the top components.
We use the \path{`docker_dependents_count'} metric as the popularity metric.

\tightpar{Transitive}
For each ecosystem, we estimate the expected number of transitive components.
First, we obtain a list of components for each ecosystem.
Then, for every component name, we use the ecosystem's default CLI (\texttt{cargo}, \texttt{go}, \texttt{mvn}, \texttt{npm}) to resolve and install the dependency in a fresh project.
A \numprint{300}~second timeout is used in this step.
After installation, the same CLI is used to list the project's dependencies and count the number of entries (omitting the project itself and the target component, if necessary).
In the end, we aggregate the dependency counts and compute the average.

\tightpar{Peer}
Building on the above, for Maven and npm, we additionally estimate the expected number of peer components.
Neither Cargo nor Go includes metadata for peer relations.

For Maven, we extract dependencies declared with the \texttt{provided}~\cite{apache_maven_dependency_mechanism_2026} scope from the target component's \texttt{pom.xml}, as this scope captures the peer relationship defined in \Cref{section:security-analysis}.
Specifically, the \texttt{provided} scope signals that the library developer will not bundle the dependency.
Instead, the deploying environment (e.g., an application server) is expected to supply it at runtime.
Consequently, whenever the declaring component is present in a deployment, its \texttt{provided} dependencies must be co-present---making them peers by definition.

For npm, we extract the \texttt{peerDependencies} from the target component's \texttt{package.json}.
This field lists the components it expects to be installed alongside, meaning that the target component is a peer of the listed components, as described in \Cref{section:security-analysis}.

For both, we ignore the fact that the peers may be optional, thus providing an upper bound.
In the end, we aggregate the dependency counts and compute the average.

\subsection{Experimental Results}
We present our experimental results obtained with zkSBOM.

\subsubsection{RQ1: End-to-End Feasibility}%

\begin{table}[t]
  \centering
  \small
  \begin{tabular}{ll r l}
    
    \toprule
    \textbf{} & \textbf{Project} & \textbf{\# Comp.}& \textbf{Incl. CVE} \\
    \midrule
    Cargo & \href{https://github.com/astral-sh/uv/releases/tag/0.9.4}{uv@0.9.4}            & 466 & CVE-2025-62518  \\
    Go    & \href{https://github.com/kubernetes/kubernetes/releases/tag/v1.29.0}{Kubernetes@1.29.0}    & 233 & CVE-2024-21626  \\
    Maven & \href{https://github.com/apache/druid/releases/tag/druid-0.22.0}{Apache Druid@0.22.0}  & 522 & CVE-2021-44228  \\
    npm   & \href{https://github.com/strapi/strapi/releases/tag/v4.4.4}{Strapi@4.4.4}        & \numprint{2308} & CVE-2023-22621  \\
    \bottomrule
  \end{tabular}%
    \caption{RQ1: End-to-end feasibility over 4 real-world projects in Cargo, Go, Maven, and npm. All subjects use CVE-2025-55182 as the non-inclusion target.}
    \label{tab:rq1-subjects}
\end{table}

We select four subjects for end-to-end evaluation, listed in \Cref{tab:rq1-subjects}.
All subjects follow the same evaluation pattern: their SBOM contains one dependency affected by a known CVE, to be included in inclusion proofs.
We use CVE-2025-55182 as the common non-inclusion proof target.
In the following, we elaborate on one subject only for sake of space, namely Apache Druid~0.22.0 in Maven, which is affected by Log4Shell (CVE-2021-44228)~\cite{noauthor_nvd_nodate} due to the usage of dependency \texttt{log4j-core~2.8.2}.

\tightpar{Inclusion proof for CVE-2021-44228}
One of the versions that CVE-2021-44228 lists as affected is \texttt{\path{log4j-core@2.8.2}}~\cite{noauthor_nvd_nodate}.
\zksbomoperator queries the committed SBOM and returns a single proof confirming \texttt{log4j-core@2.8.2} is present, which is exactly the version used as a transitive dependency in Apache Druid (third row of \Cref{tab:rq1-subjects}).
Listing~\ref{lst:inclusion} shows an excerpt of the proof output, indicating the presence of the vulnerable dependency and the cryptographically verifiable proof.
The proof is successfully verified by the zkSBOM verifier, as a \sbomconsumer would do in the field.

\begin{lstlisting}[
  caption={Abbreviated inclusion proof for CVE-2021-44228.},
  label={lst:inclusion},
  basicstyle=\ttfamily\scriptsize,
  breaklines=true,
  frame=single]
Proof: 51099f33dd...485f4f05b9
Pkg: org.apache.logging.log4j:log4j-core@2.8.2@MAVEN
\end{lstlisting}

\tightpar{Non-inclusion proof for CVE-2025-55182}
CVE-2025-55182 affects eleven package identifiers across \texttt{\path{react-server-dom-webpack}}, \texttt{\path{react-server-dom-turbopack}}, and \texttt{\path{react-server-dom-parcel}}.
Because none of these packages appear in Druid's SBOM, \zksbomoperator generates one non-inclusion proof per affected identifier, together constituting a verifiable guarantee that the artifact is not vulnerable.

Both proofs are independently verified by \texttt{zksbom-verifier} against the public commitment, confirming the full protocol feasibility: SBOM generation, SBOM upload at \zksbomoperator, vulnerability query, ZK proof generation, ZK proof verification with \zksbomverifier.

\tightpar{Other Ecosystems}
We repeat the same evaluation for the remaining three subjects in \Cref{tab:rq1-subjects}, namely Strapi for npm, Kubernetes for Go, and uv for Cargo.
In each case, \toolname successfully generates and verifies both the inclusion proof for the ecosystem-specific CVE and the non-inclusion proof for CVE-2025-55182. This strengthens the external validity of our results over four ecosystems.

\tightpar{Answer to RQ1}
\toolname is feasible end-to-end across Cargo, Go, Maven, and npm ecosystems.
Starting from real open-source projects with their SBOMs, it is able to correctly generate and verify both inclusion and non-inclusion proofs. Our experiments demonstrate that generating zero-knowledge proofs for the presence or absence of vulnerabilities is feasible for real-world software packages.

\subsubsection{RQ2: Performance}%
Beyond correctness studied in RQ1, we now study operational feasibility in terms of time and space requirements.
\begin{figure*}[t]
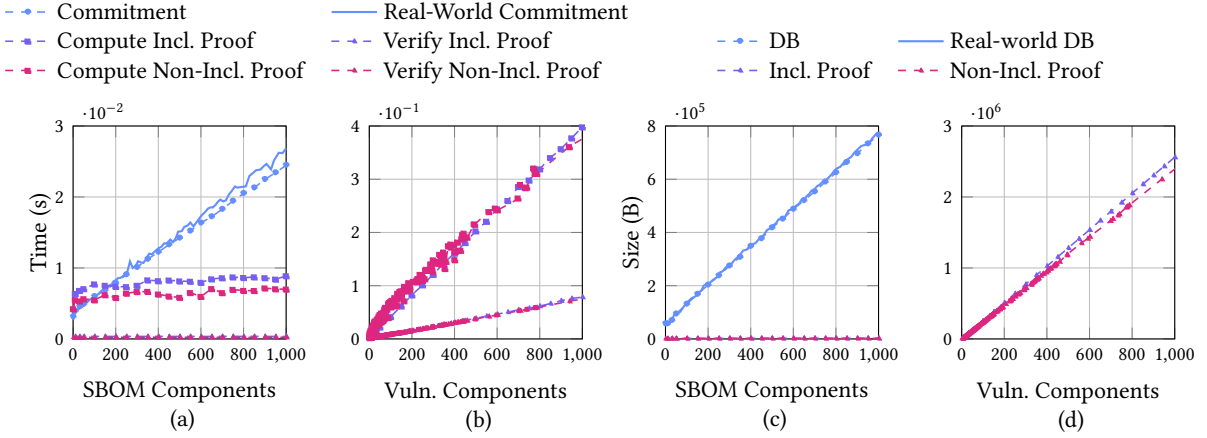

    \centering
    \begin{tikzpicture}
        \begin{groupplot}[
            group style={
                group size=4 by 1,
                horizontal sep=1.1cm, 
                x descriptions at=edge bottom,
            },
            width=4.4cm,
            height=4.4cm,           
            grid=major,
            xmin=0, xmax=1000,
            tick label style={font=\scriptsize},
            label style={font=\small},
            xlabel style={align=center}, 
        ]

            \nextgroupplot[
                ylabel={Time (s)},
                ylabel shift=-5.5pt,
                ymin=0, ymax=0.03,
                xlabel={SBOM Components \\ \vspace{0.2cm} (a)},
            ]
            \input{figures/performance-timing}

            \nextgroupplot[
                ymin=0, ymax=0.4,
                scaled y ticks={base 10:1},
                xlabel={Vuln. Components \\ \vspace{0.2cm} (b)}
            ]
           \input{figures/performance-timing2}

            \nextgroupplot[
                ylabel={Size (B)},
                ylabel shift=-5.5pt,
                ymin=0, ymax=800000,
                xlabel={SBOM Components \\ \vspace{0.2cm} (c)}
            ]
            \input{figures/performance-storage}

            \nextgroupplot[
                ymin=0, ymax=3000000,
                xlabel={Vuln. Components \\ \vspace{0.2cm} (d)}
            ]
            \input{figures/performance-storage2}
            
        \end{groupplot}

        \node[anchor=south, yshift=0.4cm] at ($(group c1r1.north)!0.5!(group c2r1.north)$) {
            \small
            \begin{tabular}{ll@{\hspace{1.5em}}ll}
                \ref{p1} Commitment & \ref{p2} Real-World Commitment \\
                \ref{p3} Compute Incl. Proof & \ref{p4} Verify Incl. Proof \\
                \ref{p5} Compute Non-Incl. Proof & \ref{p6} Verify Non-Incl. Proof \\
            \end{tabular}
        };

        \node[anchor=south, yshift=0.4cm] at ($(group c3r1.north)!0.5!(group c4r1.north)$) {
            \small
            \begin{tabular}{ll@{\hspace{1.5em}}ll}
                \ref{p7} DB & \ref{p8} Real-world DB \\
                \ref{p9} Incl. Proof & \ref{p10} Non-Incl. Proof \\
            \end{tabular}
        };

    \end{tikzpicture}
    \caption{
        RQ2: Performance analysis of \toolname.
        All graphs show our measurements for both inclusion and non-inclusion proof.
        (a)~Commitment creation time (box~(1) in~\Cref{fig:protocol}) as a function of the number of SBOM components, with the number of vulnerable components fixed at \numprint{1}.
        Also shown are proof computation and verification times (box~(4,5) in~\Cref{fig:protocol}).
        (b)~Proof computation and verification times with the number of SBOM components fixed at \numprint{1000} and a variable number of vulnerable components (box~(4,5) in~\Cref{fig:protocol}).
        (c)~Database (DB) size required to upload a single SBOM ($D$ in box~(1) in~\Cref{fig:protocol}), together with proof sizes, as a function of the number of SBOM components with the number of vulnerable components fixed at \numprint{1}.
        (d)~Proof sizes as a function of the number of vulnerable components, with the number of SBOM components fixed at \numprint{1000} (box~(6) in~\Cref{fig:protocol}).
    }
    \label{fig:performance}
\end{figure*}

\tightpar{Timing}
The blue curve in \cref{fig:performance}~(a) shows the time required to compute the commitment as the number of dependencies increases, while the number of vulnerable components is fixed.
The required execution time grows linearly.
Commitment generation for an SBOM with 0 components takes approximately 0.003~seconds, and SBOMs with \numprint{1000} components take around 0.269~seconds. 
\Cref{fig:performance}~(a) also shows that the generation time for both inclusion and non-inclusion proofs is slowly growing linearly with an increasing number of SBOM components.
Both operations require similar computation time, ranging from 0.004 to 0.009~seconds, with non-inclusion proof generation generally slightly faster.
As shown in \Cref{fig:performance}~(a), the verification time for both inclusion and non-inclusion proofs is approximately constant, while the time to verify a non-inclusion proof is slightly faster.
Both verifications require less than 0.5~ms.
\Cref{fig:performance}~(b) shows the time required with a fixed number of SBOM components and a variable number of vulnerable components.
The time to generate and verify inclusion and non-inclusion proofs both grow linearly with the number of vulnerable components, as expected, since the number of proofed components also grows linearly, and the two kinds of proofs are very similar.
Generating the proofs takes between under a millisecond for a low number of components and around 0.4~seconds for \numprint{1000} components.
The proof verification required up to 0.08~seconds for \numprint{1000} components.
These timing figures are perfectly acceptable in a real work setting.

\tightpar{Storage}
\Cref{fig:performance}~(c) shows that the size of the database files grows linearly with increasing dependency count.
The lines for synthetic and real-world SBOMs are quite similar, ranging from around 0.06~MB to 0.78~MB.
Also shown in \Cref{fig:performance}~(c), the storage required for both inclusion and non-inclusion proof files grows very slowly, with a constant number of vulnerable components and a changing number of components in the SBOM ranging from 0.001~MB to 0.003~MB.
\Cref{fig:performance}~(d) shows that with a fixed number of SBOM components and a varying number of vulnerable components, the proof sizes grow linearly and equally for inclusion and non-inclusion proofs, which is expected as the number of required proofs aligns with the number of vulnerable components.
The proof sizes reach up to 2.56~MB for \numprint{1000} vulnerable components.
These storage figures are perfectly acceptable in a real work setting.

\tightpar{Answer to RQ2}
The experimental execution times remained consistently low, ranging from less than 1~ms to a maximum of 0.4~seconds.
This near-instant performance ensures that the \sbomconsumer receives requested information without significant latency.
Storage requirements also remain minimal, with proof sizes reaching a maximum of 2.56~MB during our experimental evaluation.
These experimental results show that \toolname provides a computationally efficient and scalable solution for practical deployment of privacy-preserving SBOM sharing with \toolname.

\subsubsection{RQ3: Leakage}%
\Cref{section:security-analysis} discussed that the nature of open-source dependency trees may cause some information leakage even for a single (non-)inclusion proof.
We ran the experiment described in \Cref{sec:rq3-methodology} on April 28, 2026, and obtained the results shown in \Cref{tab:rq3}.
We used Go 1.26.2, Maven 3.6.3. npm 6.13.4, and Cargo 1.95.0 on an Ubuntu 22.04.5 system.
Due to failure to resolve components within the \numprint{300} second timeout, and some duplicate entries in the ecosyste.ms data, the total number of components is less than \numprint{10000}.

The results show that the leakage is mainly due to transitive components, with peers accounting for less than one component of leakage per (non-)inclusion proof.
If we assume unique dependencies are an order of magnitude less likely than peer dependencies, we can apply the formulas derived in \Cref{section:security-analysis} to obtain leakage estimates.
Doing this, we find that an inclusion proof leaks between
17 components (npm)
on the low end and
120 components (cargo)
on the high end.
On the other hand, a non-inclusion proof leaks approximately 1 component.

\tightpar{Answer to RQ3}
Non-inclusion proofs, which are the common case for secure and up-to-date software, leak at most one additional component.
On the other hand, an inclusion proofs my reveal a substantial number of components, depending on the ecosystem.
Note this leakage is not a result of our approach but rather a byproduct of answering the question: ``Is the software affected by this CVE?''.

These numbers are an upper bound on leakage---in practice, less leakage can be expected due to imperfect information and the use of components with private dependency information.
Moreover, the impact of the leakage depends on the overall size of an application's dependency tree.
Naturally, an ecosystem with a higher number of transitive components, such as Cargo, will result in larger overall dependency trees.
Unfortunately, no data is available on application sizes of different ecosystems, so we cannot provide an estimate.

\begingroup
\setlength{\tabcolsep}{0.95\tabcolsep}
\begin{table}[t]
  \centering
  \label{tab:rq3}
  \small
  \begin{tabular}{@{} l r r r r r r r @{}}
    \toprule
    & \multicolumn{1}{l}{\multirow{2}{*}{\textbf{Direct}}} & \multicolumn{2}{c}{\textbf{Transitive}} & \multicolumn{2}{c}{\textbf{Peer}} & \multicolumn{2}{c}{\textbf{Leakage}} \\
    \cmidrule(lr){3-4} \cmidrule(lr){5-6} \cmidrule(lr){7-8}
    & & \textbf{Total} & \textbf{Avg.} & \textbf{Total} & \textbf{Avg.} & \textbf{Inc.} & \textbf{Exc.} \\
    \midrule
    Cargo  & \numprint{8613} & 1.03M & 120.10 & -    & -    & 120.10 & 1.21 \\
    Go     & \numprint{8792} & 0.45M &  51.21 & -    & -    &  51.21 & 0.52 \\
    Maven  & \numprint{6209} & 0.13M &  21.38 & 4.5k & 0.65 &  35.78 & 1.00 \\
    npm    & \numprint{7242} & 0.10M &  13.78 & 1.8k & 0.24 &  17.29 & 0.42 \\
    \bottomrule 
  \end{tabular}%
  \caption{
    RQ3: Leakage results.
    \emph{Direct} is the number of unique components considered for the \emph{Transitive} and \emph{Peer} results.
    \emph{Leakage} is the expected leakage resulting from one inclusion (inc.) or non-inclusion (exc.) proof. %
  }
\end{table}
\endgroup

\section{Discussion} \label{section:discussion}

\subsection{Threats to Validity} \label{section:threats-to-validity}
\tightpar{Internal validity}
A bug in \toolname may invalidate the results of RQ1.
For RQ2, the use of synthetic data may lead to representativeness issues.
We account for this by, where possible, comparing synthetic data results to a real-world benchmark and observing near-identical results.

\tightpar{External validity}
The results of RQ1 hold over four real-world projects.
All measurements of RQ2 were conducted on a single laptop, and performance may differ on containerized or server-side machines or on different hardware architectures.
The results of RQ3 are skewed towards popular packages and may not generalize to the entire ecosystem.

\subsection{Limitations} \label{section:limitations}

\tightpar{SBOM accuracy}
\toolname does not address the problem of SBOM accuracy~\cite{balliu2023challenges}.
Consequently, an artifact may be incorrectly proven safe when a vulnerable dependency is missing, or vulnerable due to a phantom dependency (e.g., an unused component).
Also, a mismatch in component identifiers between the SBOM and \securitydb can result in incorrect proofs.

\tightpar{SBOM truthfulness}
Per our threat model, we assume that \softwaresupplier generates SBOMs honestly.
Addressing this limitation can be addressed by attestation-based approaches~\cite{hugenroth2025attestable,kim_attestation-based_2025} or using verifiable execution with zkVMs for SBOM generation.

\section{Related Work}
We first discuss closely related work, then place our work in the broader context of supply chain security.

\tightpar{Privacy-preserving SBOMs}
Ishgair et al.\ propose \name{Petra}~\cite{ishgair2025trustworthy}, an SBOM exchange system using Ciphertext-Policy Attribute-Based Encryption (CP-ABE) to selectively encrypt nodes within an SBOM.
The result is that only stakeholders can access the parts of the SBOM they are authorized to read.
\name{Petra} and \toolname share the same motivation but use fundamentally different methods to achieve privacy-preserving SBOM sharing.
In contrast to \toolname, \name{Petra} reveals component use regardless of the component state, placing high trust in the stakeholders with access.
Our paper uniquely provides a systematic evaluation of integrity properties for both the \sbomconsumer and \softwaresupplier.

A concurrent preprint by Castiglione et al.\ proposes \name{VeriSBOM}~\cite{castiglione2026verisbom}, a framework for privacy-preserving SBOM sharing based on vector commitments and Zero-Knowledge Succinct Non-interactive Arguments of Knowledge (zkSNARKs). 
The system requires a maintained, public (sparse) Merkle tree of components and an isomorphic shadow tree that embeds security properties verified by an auditor.
This enables \softwaresupplier{s} to include zero-knowledge proofs of these trees in their SBOM, providing verifiable attestations of security-relevant statements for stakeholders. 
In contrast to \toolname, \name{VeriSBOM}'s dual-tree architecture requires cooperation with an additional third party, complicating real-world deployment.
More importantly, \name{VeriSBOM} does not handle security properties that change over time, such as the presence of a known vulnerability.
In terms of performance, \toolname has lower proving complexity than \name{VeriSBOM} due to the use of ZKS.

Compared to these works, \toolname is the first to identify and analyze the leakage problem that occurs when privacy-preserving proofs are used on top of public ecosystem data.

Work by Deng et al.\ \cite{dengzero} outlines a conceptual ZKP framework designed to verify SBOM security properties, such as the absence of critical vulnerabilities, while maintaining confidentiality of sensitive supply chain data.

\tightpar{Cryptography for Supply Chain Security}
Digital signatures are a widely deployed approach for linking a software artifact to a \softwaresupplier, with solutions like in-toto~\cite{torres2019toto} targeting open source ecosystems specifically.
However, they do not support the use case of privacy-aware SBOM sharing and vulnerability querying.

Private Set Intersection (PSI) protocols~\cite{morales2023private}, Secure Multi-Party Computation (SMPC)~\cite{ZHAO2019357}, and Trusted Execution Environments (TEEs) ~\cite{hugenroth2025attestable,kim_attestation-based_2025} are possible implementation paths to privacy-aware SBOM sharing, but, to the best of our knowledge, nobody has studied them in the context of SBOM sharing. Our approach is the first end-to-end implemented and systematically analyzed blueprint architecture.

\name{Cheesecloth}~\cite{cuellar2023cheesecloth} employs zkSNARKs for proving properties about software vulnerabilities, focused on proving the existence of bugs in programs rather than querying for vulnerabilities. Our paper is the first to apply zero-knowledge sets to the specific problem of SBOM sharing.

\section{Conclusion}
We presented \toolname, a novel privacy-preserving SBOM sharing protocol.
Based on zero-knowledge sets, \toolname allows \sbomconsumer{s} to verify whether a software artifact is affected by a known vulnerability without requiring direct access to the \softwaresupplier's SBOM.
We have defined a threat model for SBOM sharing and conducted a systematic security analysis of \toolname. 

\begingroup
\parfillskip=0pt
We showed \toolname meets all security requirements,
we quantified leakage attributable to inclusion and non-inclusion proofs given knowledge of public data.
We validated our approach through a publicly available prototype implementation, tested it across four popular package ecosystems, and evaluated it on over \numprint{43000} real-world SBOMs.
Our results show that \toolname is secure, computationally efficient, and usable on large, real-world software applications. 
\toolname offers a sound and practical solution for privacy-aware SBOM sharing, addressing the conflicting needs of software transparency and software confidentiality.
\par
\endgroup

\ifcameraready%
\begin{acks}
This work was supported by the Swedish Foundation for Strategic Research (SSF), the Swedish Research Council (VR), Digital Futures, and Wallenberg AI, Autonomous Systems and Software Program (WASP) funded by the Knut and Alice Wallenberg Foundation.
\end{acks}
\fi

\bibliographystyle{ACM-Reference-Format}
\bibliography{sample-base}

\appendix

\section{Open Science}
\toolname artifacts are open source under the MIT License and are available at \url{https://github.com/chains-project/zkSBOM}.

The repository is structured as follows:
\texttt{zksbom-operator} and \texttt{zksbom-verifier} contain the implementations of \zksbomoperator and \zksbomverifier, respectively.
The code used to produce the results for each RQ is located in its corresponding \texttt{rqX} directory.
Additionally, GitHub Actions workflows used for RQ1 and RQ3 are provided in \texttt{\path{zksbom/.github/workflows}}.

\section{Ethical Considerations}
The focus of \toolname is to increase the security when sharing SBOMs.
It focuses on balancing security and privacy, trying to protect a \softwaresupplier's intellectual property as well as reducing the attack surface in the context of SBOM sharing as much as possible.
We therefore do not have any ethical considerations in allowing \softwaresupplier to selectively share their SBOMs with \sbomconsumer through such a system, in a more privacy-preserving manner.

To ensure \toolname aligns with real-world challenges and requirements, we engaged with industry stakeholders, which helped us refine the system.
We believe \toolname is a solution that balances transparency and security.

\section{Generative AI Usage}
Claude Code (Anthropic) and Cursor were used for writing assistance throughout this paper, with the following scope.

\tightpar{Research Questions}
For the RQ1 results section (\Cref{section:evaluation}), including the written description of the inclusion and non-inclusion proof outputs and the ``Answer to RQ1'' paragraph, the initial prose was drafted by the AI tools based on the experimental outputs and data.
Similarly, the descriptive text accompanying \Cref{tab:rq1-subjects} was initially drafted with AI assistance.
However, the first draft has been reviewed and further refined by all authors.
For RQ3 experiments, we used Claude Code for generating scripts.
The execution and output of the scripts are not influenced by AI.
Nonetheless, the generated code is also reviewed by the authors.

\tightpar{Editorial refinement}
All remaining sections (excluding the bibliography) were edited using Claude Code and Cursor for grammar, clarity, and style.
Grammar was additionally checked using Grammarly.

\tightpar{Author validation}
All AI-generated and AI-assisted text was carefully reviewed, revised where necessary, and approved by all authors.
The authors take full responsibility for the content of this work.

\tightpar{Code generation}
The development of \toolname made partial use of Claude Code, an AI-assisted coding tool.
Specifically, AI assistance was used for the C++ FFI layer in \zksbomoperator and \zksbomverifier, and for implementing scripts in RQ2.
All generated artifacts were thoroughly reviewed and tested by the authors to ensure correctness.

\end{document}